# Reply to "Comment on 'Optical detection of transverse spin-Seebeck effect in permalloy film using Sagnac interferometer microscopy' "

R. McLaughlin[1], D. Sun[1,2], C. Zhang[1,3], M. Groesbeck[1], and Z. Valy Vardeny[1]

[1]*Department of Physics & Astronomy, University of Utah, Salt Lake City, Utah, 84112, USA*
[2]*Department of Physics, North Carolina State University, Raleigh, NC, 27695, USA*
[3]*Institute of Chemistry, Chinese Academy of Science, Beijing, 10019, China.*

*To whom correspondence should be addressed: val@physics.utah.edu

In their Comment [1], Kimling and Kuschel (hereafter the 'commenters') challenge our original interpretation of the magneto-optical Kerr effect (MOKE) measurements using ultrasensitive Sagnac interferometer [2], claiming that the transverse spin-Seebeck effect (TSSE) is not the only contribution to the measured change in the Kerr rotation angle from a $Ni_{80}Fe_{20}$ (NiFe) ferromagnetic (FM) slab subjected to a lateral temperature gradient in the presence of a magnetic field. The authors assert that: (1) the TSSE, in general, has not been completely proven so far, and that the existing theories, in particular the 'phonon magnon drag' model that we have used in our publication, in fact cannot explain the original work of Uchida *et al* [3]. (2) The commenters based their critique on an estimate of the magnitude of the measured effect, arguing that to observe the TSSE that we originally claimed, the temperature gradient in our measurements should have been much larger than the value we measured.

In this Reply, firstly, for the reader sake, we summarize previous literature reports on the TSSE response in FM metallic systems. Secondly, we dispute the estimate of the necessary large temperature gradient made by the commenters. Importantly, the Kerr effect sensitivity to spin accumulation (via the TSSE response) has not been recognized by the commenters; in fact it was mistakenly assumed that the Kerr effect sensitivity to the spin accumulation by the TSSE is the same as the Kerr sensitivity to the bulk magnetization change due to the temperature change. In conclusion, we show that the TSSE is the only viable interpretation of our original measurements, and that the 'phonon-magnon drag model' is indeed capable of explaining our results.

## 1. Introduction

The elusive phonon-mediated transverse spin Seebeck effect (TSSE) in FM slabs has been previously supported by the so-called 'patterned configuration' [4,5], where the FM slab is divided into thin, isolated stripes. In this case, the spatially distributed SSE may still exist, and, due to the stripe isolation, without any interference of other spurious artefacts such as the spin-dependent Seebeck effect (SDSE). Such measurements were performed before in NiFe [4] and GaMnAs [5] using electrical-based detection, namely via the inverse spin Hall effect (ISHE) in a nonmagnetic overlayer with large spin-orbit coupling, namely Pt. This type of measurement has been considered by many to be a direct evidence of TSSE response, since only TSSE would be generated by the long-range phonons propagation through the substrate via the magnon-phonon coupling (i.e., 'phonon-magnon drag model').

We note that the TSSE response measured via electrical-based ISHE detection is still under intense debate, due to the possible interference from various thermoelectric and magneto-thermoelectric artifacts. Specifically when the ISHE is used to measure the TSSE in a FM slab, the additional top contacts might influence the thermal conductivity of the system, and consequently induce an unintentional vertical temperature gradient along the FM/Pt interface that leads to overwhelming artefact signals arising from the magneto-thermoelectrics effect. *This is an unfortunate disadvantage of the electrical detection measurement.* Thus all arguments against the existence of TSSE response in FM slabs have been based on the same type of electrical measurement, namely via ISHE. This suggests that a 'cleaner' and artefact-free detection scheme is required to provide an alternative route of proving and studying the TSSE response. In this respect our optical detection scheme excludes all of the electrical artefacts. We thus believe that the TSSE response may be subtle when trying to measure it via an electrical detection scheme, *but it might be clearly observable via magneto-optical detection* [2].

## 2. Comparison of Kerr sensitivity to spin accumulation via TSSE and bulk magnetization via 'thermal effect'

The commenters claimed that for observing the TSSE response via magneto-optic measurements the temperature gradient in the FM slab should have been much larger than the measured value. One key parameter used in the commenters' equation to estimate the temperature gradient is the Kerr sensitivity to spin accumulation via TSSE. For this estimate the commenters used an incorrect sensitivity parameter that was taken from the original S.I. figures (Figs. S6 and S7), which in fact characterizes the Kerr sensitivity parameter to the reduction of magnetization due to the temperature change upon uniform heating, rather than the Kerr sensitivity to spin accumulation on the FM slab surface. These are **two different types** of Kerr sensitivity, which are orders of magnitude different, and thus should not be confused. Figure 1 clearly demonstrates the difference between the two Kerr sensitivities. Figures 1(a) to 1(c) illustrate the Kerr signal response along the slab direction (x) under three different conditions, when a **small temperature gradient** is applied to the FM slab. These are: 1(a); heating induced decrease in the FM magnetization, which exhibits a progressive Kerr angle reduction along x, as expected. 1(b); spin accumulation via TSSE which induces an opposite Kerr signal at each end of the slab. Here the Kerr signal increases at the left end but decreases at the right end, having the same amplitude. (c) The measured Kerr signal in the actual experiment, where both Kerr responses (due to heat and TSSE) contribute together. In this case the contribution to the Kerr signal due to heat may be derived from the asymmetry in the Kerr signals at the two ends of the FM slab, *which is less than 5% at low temperature gradient* similar as in our original paper [2].

Figures 1(c) to 1(f) show the Kerr(x) response in the same FM slab **at large applied temperature gradient**. In this case, the temperature of the left end can no longer be maintained at room temperature (see Fig. 1d). This leads to a reduction of magnetization (and corresponding Kerr signal) at both ends of the slab. Note that the slope in Fig. 1(d) is larger than that in Fig. 1(a) due to the large temperature gradient. Meanwhile, the Kerr signal due to the TSSE response also increases. However by combining the two Kerr contributions as shown in Fig. 1(f) it is no longer possible to distinguish between the Kerr response due to heating and that related to the TSSE. The Kerr sensitivity to the TSSE claimed by the commenters was probably extracted from this situation, and therefore is erroneous. We note that all the optical TSSE coefficients extracted in our original paper were determined only at the small-temperature gradient condition [2].

We now estimate the correct Kerr sensitivity to the heat and TSSE effects from our original data. For this purpose we analyze in more detail the dynamics and magnitude of the Kerr response at the left end, $\Delta\theta_L(t)$, and right end, $\Delta\theta_R(t)$ upon applying a temperature gradient, that were depicted in Fig. 2 of our original paper. We note that at small temperature gradient $\Delta\theta_R(t)$ saturation value is somewhat larger than that of $\Delta\theta_L(t)$. This can be readily explained if there is an additional contribution to $\Delta\theta_R(t)$ that is not related to that of TSSE. We identify the additional contribution to the dcrease in $\theta_R(t)$ as due to a 'thermal effect', as shown in Fig. 1 in this Reply, where the FM magnetization of the right end decreases when its temperature increases. A crude estimate shows that this thermal contribution to $\Delta\theta_R(t)$ is about 5% of the total TSSE-related $\Delta\theta$ at saturation; namely $[|\Delta\theta_R| -\Delta\theta_L]/[ |\Delta\theta_R| +\Delta\theta_L]=5\%$. This shows that in NiFe the Kerr angle sensitivity associated with the TSSE is at least one order of magnitude larger than that associated with the change in magnetization due to the temperature increase, at the conditions set in the experiment, i.e. temperature difference, *ΔT*=2.0 K between the left and right ends of the FM slab. This argument, based on the experimental results in our original paper refutes the assumption made in the Comment that the TSSE sensitivity to MOKE is the same as that of the bulk magnetization. We speculate that the higher MOKE sensitivity to the TSSE is the accumulation of the excess spins at the upper surface of the NiFe stripe (or film) within the optical skin depth of the laser excitation beam, compare to the spin density reduction in the bulk of the FM film. This observation calls into question the exaggerated assumption used for the *ΔT* estimate made in the Comment.

In summary, we conclude that the MOKE sensitivity to spin accumulation on the NiFe surface vastly exceeds that of the magnetization variation in the NiFe bulk, and therefore is possible to measure TSSE by optical means.

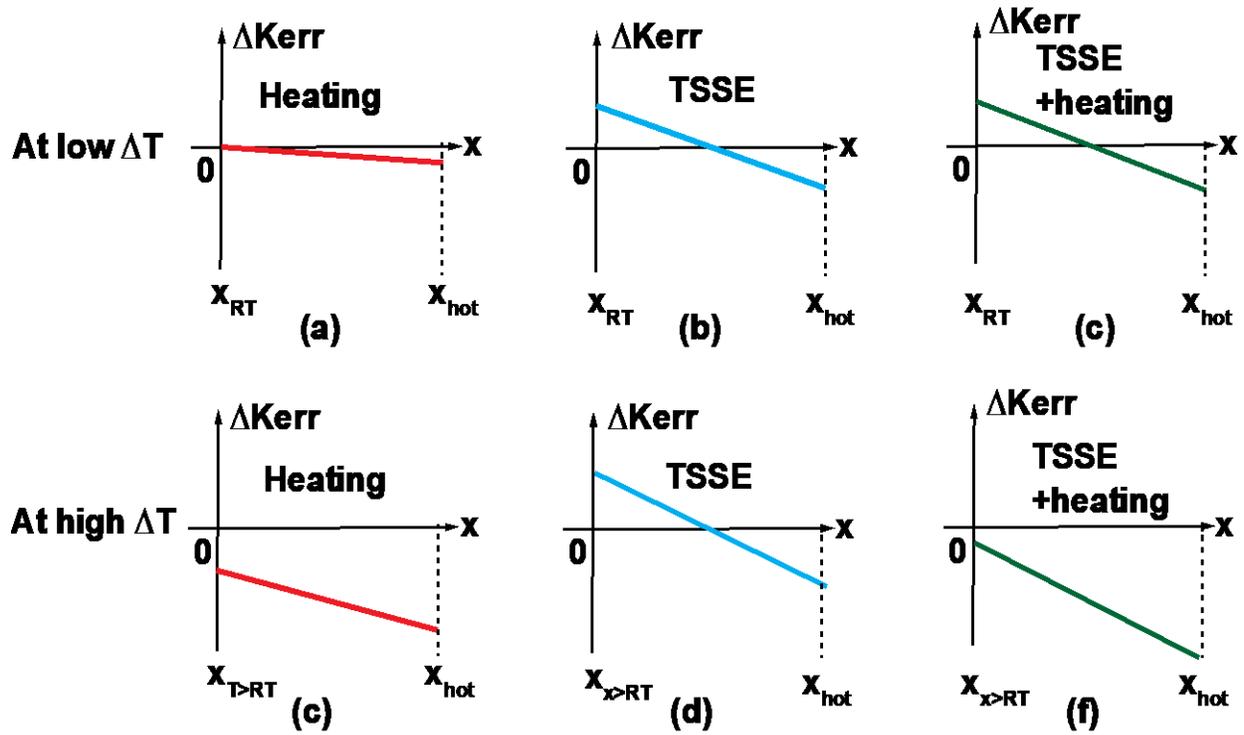

*Figure 1 Schematic illustrations of the change, ΔKerr(x) response in the Kerr angle in a NiFe slab upon heating its right end. The x-axis represents the position along the FM slab. Panels (a)-(c) are for small applied temperature difference, ΔT between the slab ends; whereas panels (d)-(f) are for large ΔT. Panels (a) to (c) [and panels (d) to (f)] describe, respectively ΔKerr(x) upon heating, TSSE, and their combined effect that occurs in the measurement.*

We thank Dr H. Liu for the help in performing the TSSE experiment on a patterned NiFe film. The work at the University of Utah was supported by MURI-AFOSR grant FA9550-14-1-0037 (TSSE on NiFe slab) and by NSF-EAGER grant DMR-1836989 (TSSE analysis). D.S. was supported by a start-up grant from the North Carolina State University.

**References:**

1. J. Kimling and T. Kuschel, Comment on "Optical detection of transverse spin-Seebeck effect in permalloy film using Sagnac interferometer microscopy".
2. R. McLaughlin, D. Sun, C. Zhang, M. Groesbeck, and Z.V. Vardeny, Optical detection of transverse spin-Seebeck effect in permalloy film using Sagnac interferometer microscopy. Phys. Rev. B **95**, 180401 (2017).
3. K. Uchida, S. Takahashi, K. Harii, J. Ieda, W. Koshibae, K. Ando, S. Maekawa, and E. Saitoh, Observation of the spin Seebeck effect. Nature **455**, 778 (2008).
4. K. Uchida, H. Adachi, T. An, T. Ota, M. Toda, B. Hillebrands, S. Maekawa, and E. Saitoh, , Long-range spin Seebeck effect and acoustic spin pumping, Nat. Mater. **10**, 737 (2011).
5. C.M. Jaworski, J. Yang, S. Mack, D.D. Awschalom, J.P. Heremans, and R.C. Myers, Observation of the spin-Seebeck effect in a ferromagnetic semiconductor. Nat. Mater. **9**, 898 (2010).